\documentclass[Physsubmission, Phys]{SciPost}

\hypersetup{
    colorlinks,
    linkcolor={red!50!black},
    citecolor={blue!50!black},
    urlcolor={blue!80!black}
}

\usepackage[bitstream-charter]{mathdesign}
\usepackage{subfig}
\usepackage{graphicx}
\DeclareSymbolFont{usualmathcal}{OMS}{cmsy}{m}{n}
\DeclareSymbolFontAlphabet{\mathcal}{usualmathcal}

\begin{document}

\begin{center}{\Large \textbf{
Precision measurements of jet and photon production at the ATLAS experiment \footnote{Copyright 2022 CERN for the benefit of the ATLAS Collaboration.
CC-BY-4.0 license.} \\
}}\end{center}

\begin{center}
Giuseppe Callea on behalf of the ATLAS collaboration\textsuperscript{1$\star$}
\end{center}

\begin{center}
{\bf 1} University of Glasgow
\\
* giuseppe.callea@cern.ch
\\
\end{center}

\begin{center}
\today
\end{center}


\definecolor{palegray}{gray}{0.95}
\begin{center}
\colorbox{palegray}{
  \begin{tabular}{rr}
  \begin{minipage}{0.1\textwidth}
    \includegraphics[width=23mm]{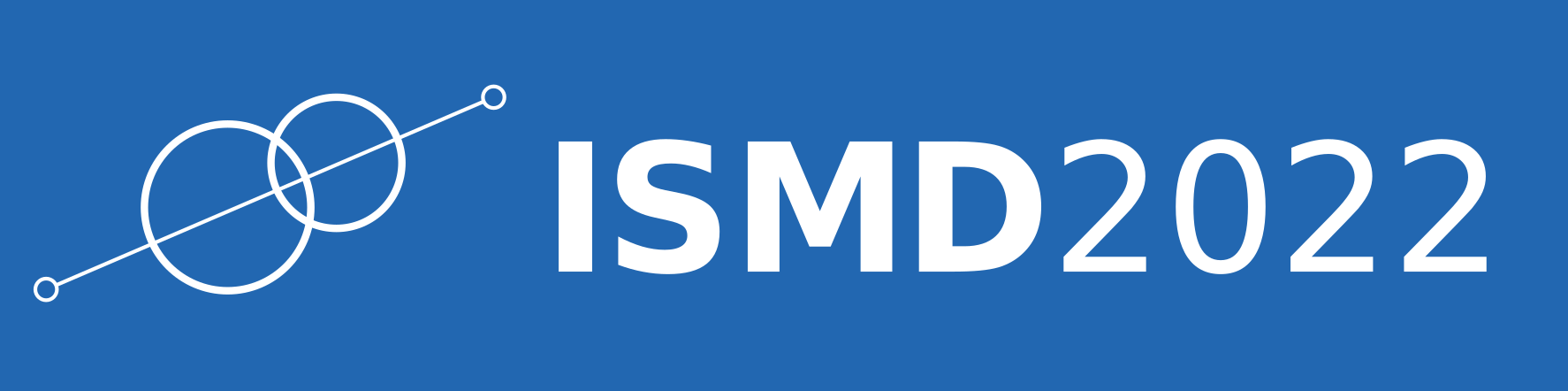}
  \end{minipage}
  &
  \begin{minipage}{0.8\textwidth}
    \begin{center}
    {\it 51st International Symposium on Multiparticle Dynamics (ISMD2022)}\\ 
    {\it Pitlochry, Scottish Highlands, 1-5 August 2022} \\
    \doi{10.21468/SciPostPhysProc.?}\\
    \end{center}
  \end{minipage}
\end{tabular}
}
\end{center}

\section*{Abstract}
{\bf
The production of jets and prompt isolated photons at hadron colliders provides stringent tests of perturbative QCD. We present the latest measurements of photon+jets and diphoton production using proton-proton collision data collected by the ATLAS experiment at the LHC. The measurements are compared to state-of-the-art NLO and NNLO predictions.
}

\section{Introduction}
\label{sec:intro}
The production of prompt photons in association with hadronic jets in proton-proton collisions, $pp \to \gamma + \mathrm{jet} + X$, provides a testing ground for perturbative Quantum Chromodynamics (pQCD) in a cleaner enviroment than in jet production, since the colourless photon originates directly from the hard interaction. The measurement of the angular correlations between the photon and the jet can be used to probe the dynamics of the hard-scattering process. Diphoton production offers a clean final state for the study of the properties of the Higgs boson and a possible window into new physics phenomena. Prompt photon production can proceed via `direct' production, where the photon originates from the hard scatter, or the `fragmentation' process where it is produced in the fragmentation of a parton with high $p_\mathrm{T}$~\cite{DirectFrag}. 
This contribution summarises the studies on the photon + two jet~\cite{gamma2jets} and photon pair productions~\cite{diphoton} with the ATLAS detector~\cite{ATLAS} at the Large Hadron Collider (LHC). 

\section{Photon + 2 jets production}
\label{sec:gamma2jets}
\begin{figure}[!htbp]
\centering
\subfloat[]
{\includegraphics[width=0.4\textwidth]{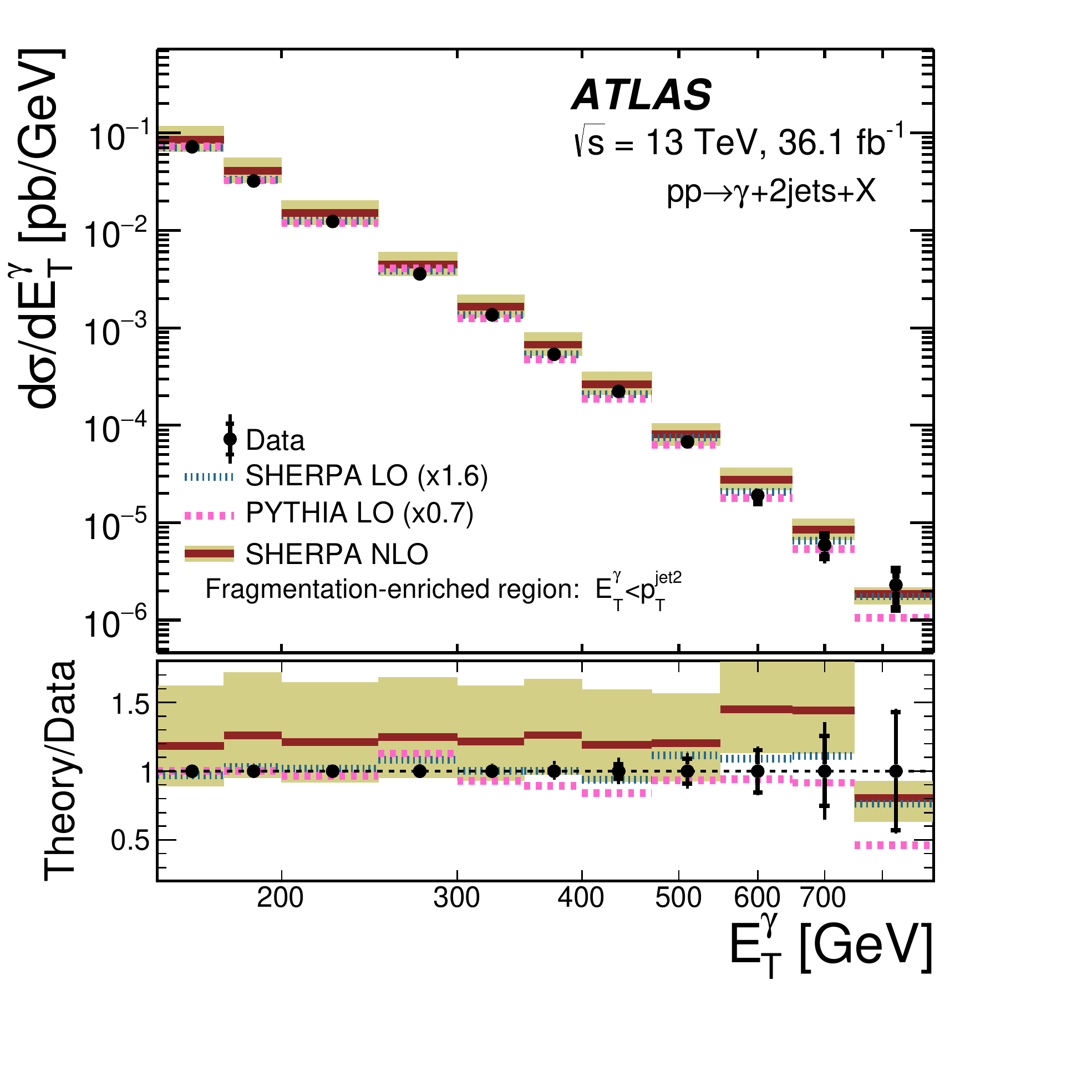}
\label{fig:Etgamma}} 
\subfloat[]
{\includegraphics[width=0.4\textwidth]{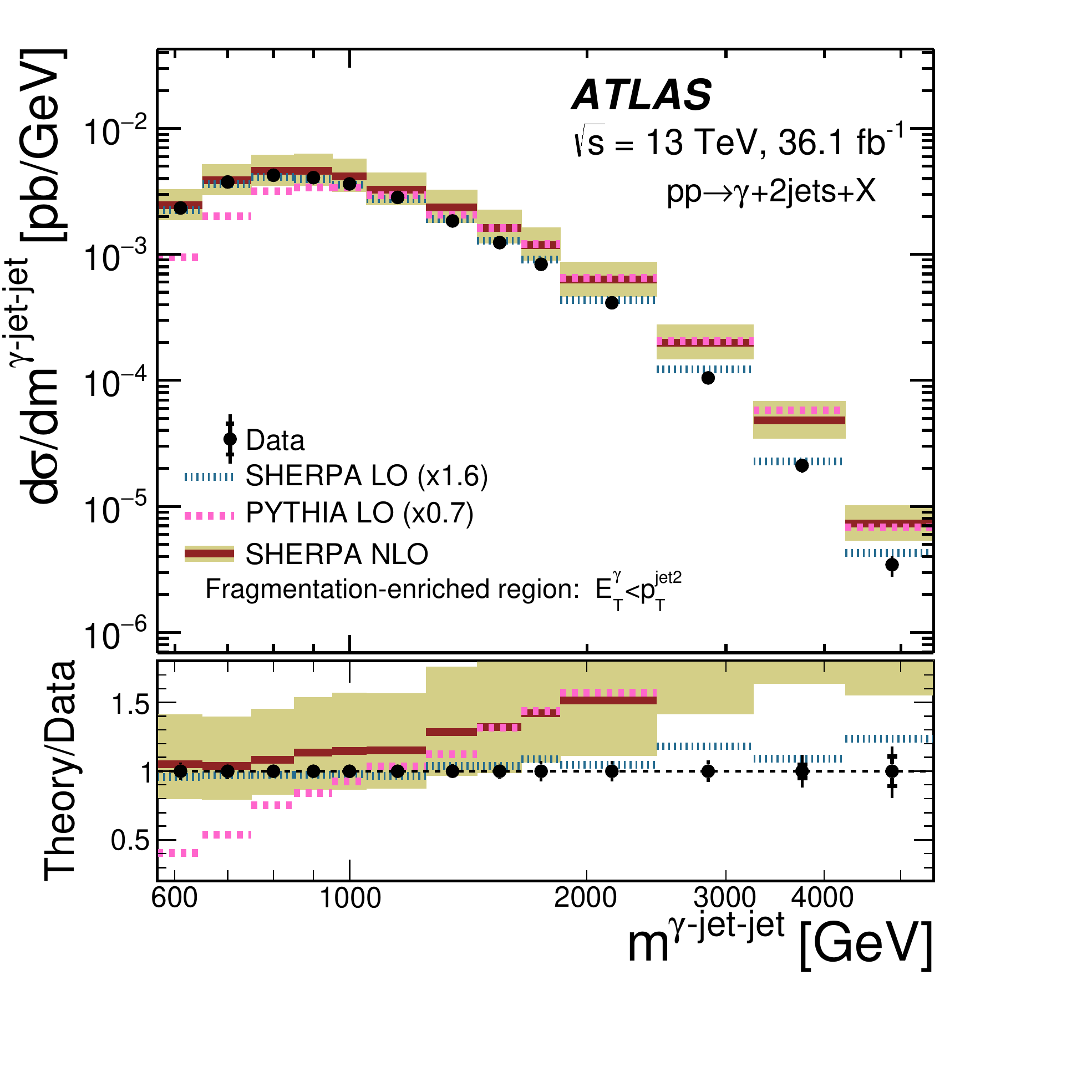}
\label{fig:combmass}}
\caption[]{
Differential cross-sections of the photon + 2 jet production measured as a function of Figure $E_\mathrm{T}^{\gamma}$~(a) and $m_{\gamma\textrm{-}\mathrm{jet}\textrm{-}\mathrm{jet}}$~(b) for both direct and fragmentation process and compared to the theoretical MC predictions~\cite{gamma2jets}.
\label{fig:photonplus2jetsplot}}
\end{figure}

The production of prompt photons with two associated jets provides a solid testing ground for pQCD by looking at the angular correlations between the photon and each of the jets and between the two jets. In addition, measurements of the invariant-mass distributions of the dijet system and the $\gamma$-jet-jet system are sensitive to the dynamics of the hard interaction. A comprehensive study of the observables describing this final state is of relevance for the development of pQCD calculations as well as for the tuning of Monte Carlo (MC) models. These studies were performed using a dataset with an integrated luminosity of 36.1 fb$^{-1}$, collected with the ATLAS detector during Run 2. \\
Photons are required to pass isolation and identification requirements and have $E_\mathrm{T}^{\gamma}$ > 150 GeV. Jets are reconstructed with the anti-$k_t$ algorithm with radius parameter R=0.4 and are required to have $p_\mathrm{T}$ > 100 GeV. A `fragmentation-enriched' sample is selected by requiring $E_\mathrm{T}^{\gamma}$ < $p_\mathrm{T}^{\mathrm{jet}2}$, where $p_\mathrm{T}^{\mathrm{jet}2}$ is the transverse momentum of  the sub-leading jet. A `direct-enriched' sample is selected with the requirement $E_\mathrm{T}^{\gamma}$ > $p_\mathrm{T}^{\mathrm{jet}1}$, where $p_\mathrm{T}^{\mathrm{jet}1}$ is the $p_\mathrm{T}$ of the leading jet. The tree-level plus parton-shower predictions from leading order (LO) Sherpa~\cite{Sherpa} and Pythia8~\cite{Pythia} as well as the next-to-leading-order (NLO) QCD predictions from Sherpa are compared with the measurements. Figure~\ref{fig:photonplus2jetsplot} shows the differential cross-section of the photon + 2 jet production as a function of $E_\mathrm{T}^{\gamma}$ and the invariant mass of the $\gamma$-jet-jet system, $m_{\gamma\textrm{-}\mathrm{jet}\textrm{-}\mathrm{jet}}$ . The NLO QCD predictions of Sherpa describe the data adequately in shape and normalisation except for regions of phase space such as those with high values of the invariant mass, where the predictions overestimate the data. The Sherpa LO predictions are in good agreement with the data. 

\section{Diphoton production}
\label{sec:diphoton}
\begin{figure}[!h]
\centering
\subfloat[]
{\includegraphics[width=0.36\textwidth]{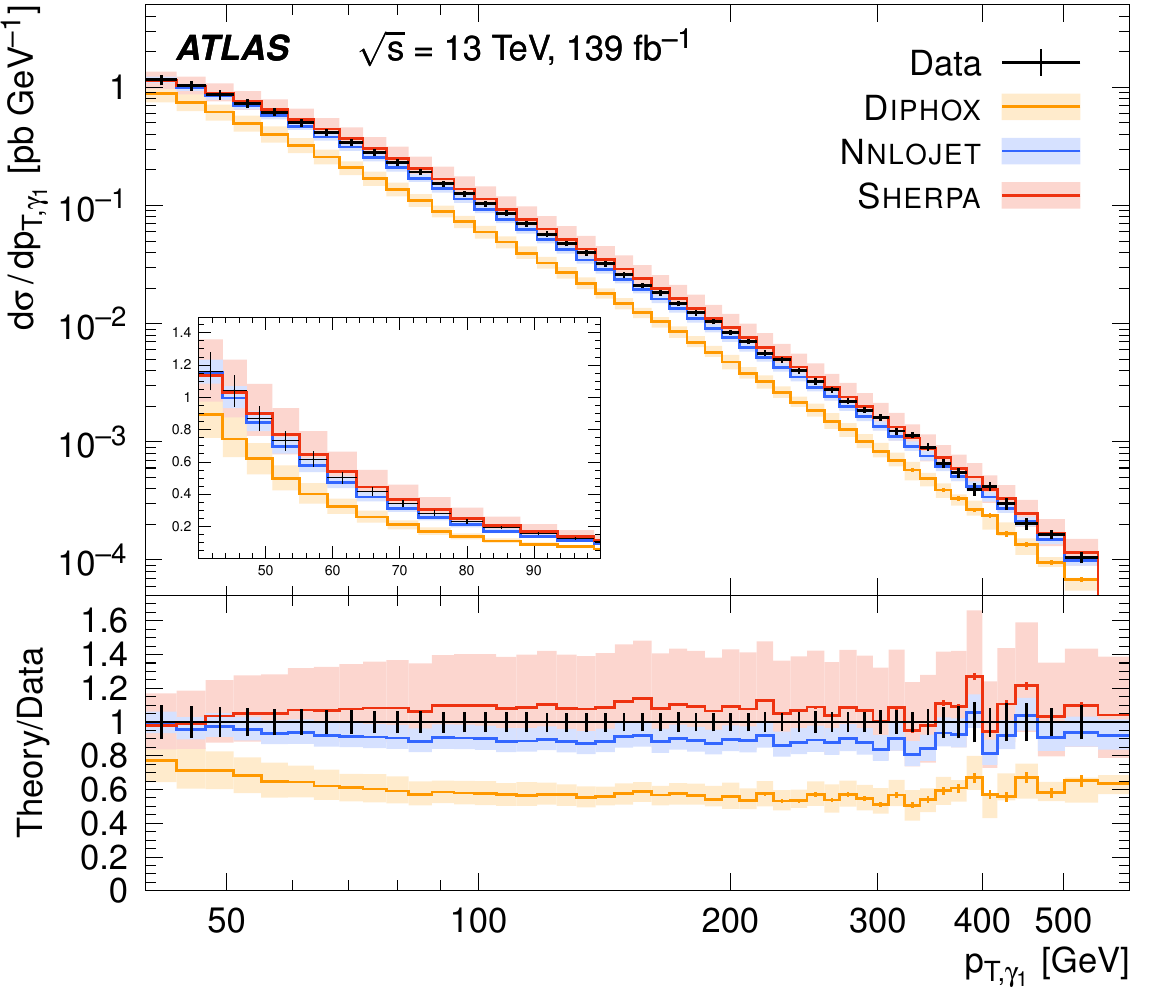}
\label{fig:ptgammagamma}} 
\subfloat[]
{\includegraphics[width=0.36\textwidth]{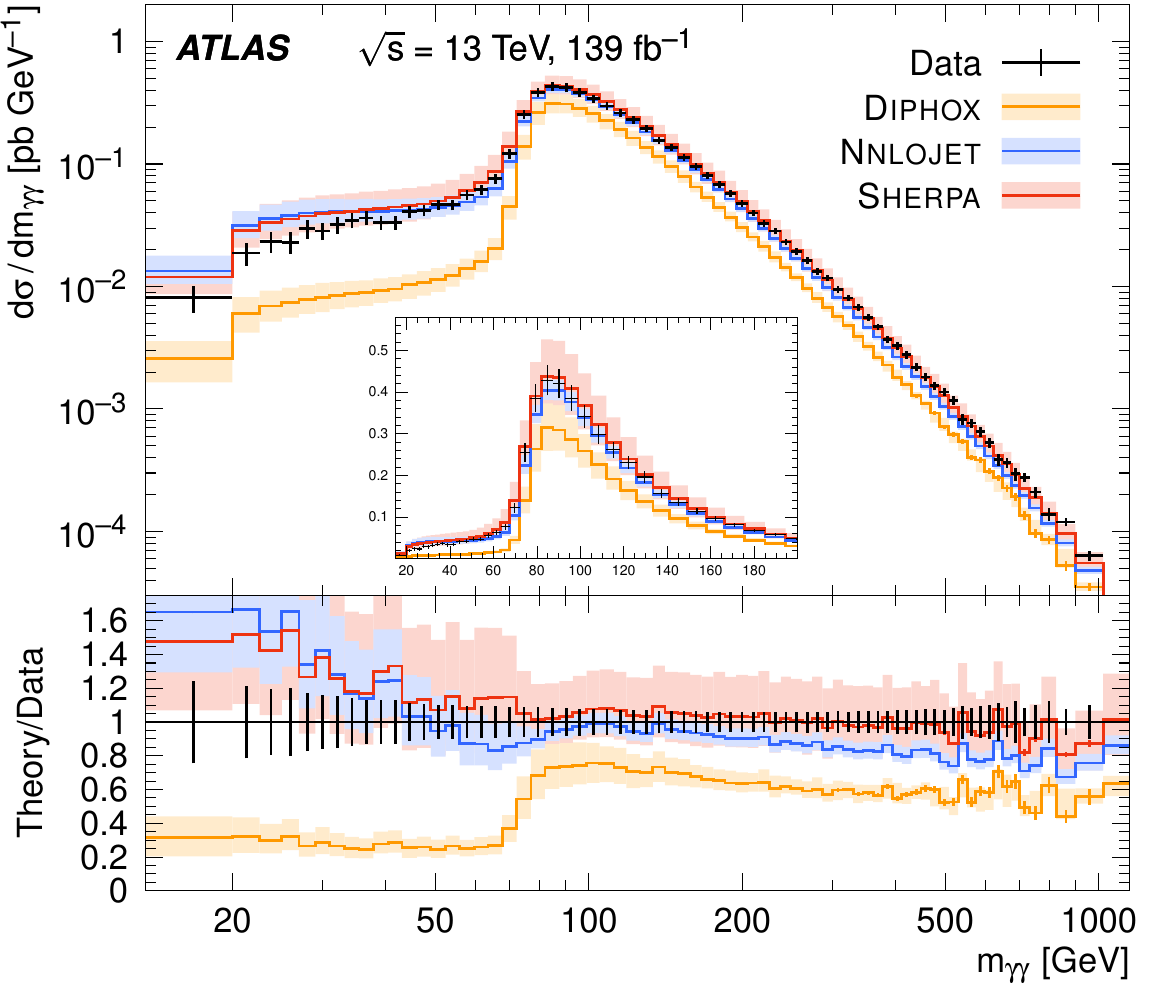}
\label{fig:diphotonmass}}
\subfloat[]
{\includegraphics[width=0.32\textwidth]{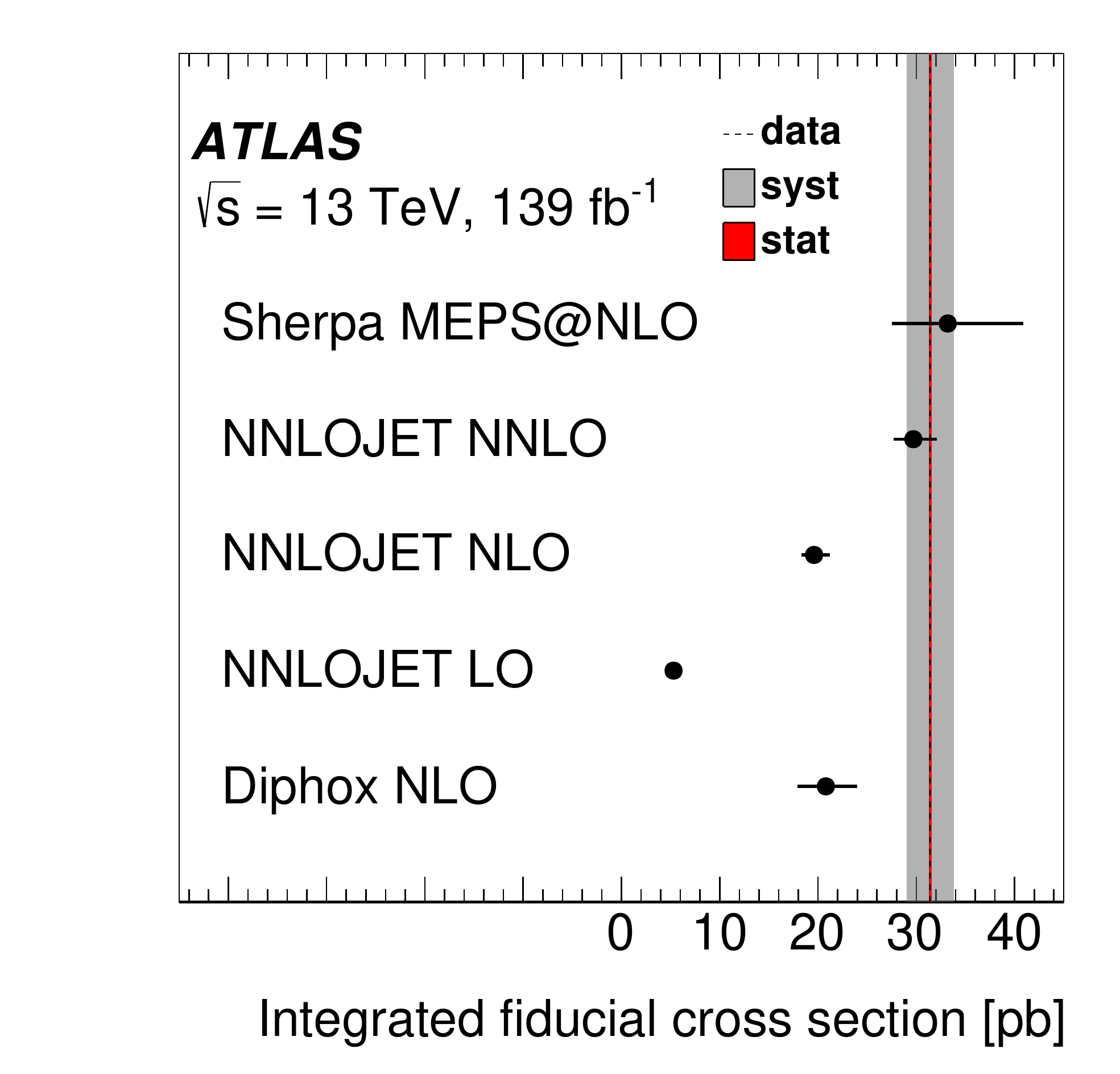}
\label{fig:integratedxsec}}
\caption[]{
Differential cross-sections of the photon pair production measured as a function of $p_\mathrm{T}^{\gamma,1}$~(a) and $m_{\gamma\gamma}$~(b) for both direct and fragmentation processes and compared to the theoretical MC preditions. Figure~(c) shows the integrated fiducial cross section comparisons between the ATLAS data and the theoretical predictions~\cite{diphoton}.
\label{fig:diphotonplots}}
\end{figure}

Measurements of a photon pair production were studied with the full Run 2 dataset collected with the ATLAS detector with an integrated luminosity of 139 fb$^{-1}$. At least two photon candidates are required per event with transverse momentum above 40 GeV and 30 GeV, respectively. The photon pseudorapidity is required to be $|\eta^{\gamma}|$ < 2.37, excluding the transition region between the barrel and the endcap (1.37 < $|\eta^{\gamma}|$ < 1.52). In this acceptance region, the high granularity of the calorimeter system allows the efficient identification of photons. The unfolded data were compared to several state-of-the-art predictions: Fixed-order NNLO with NNLOJet framework~\cite{nnlojet} to obtain the diphoton predictions up to NNLO QCD accuracy; Fixed-order NLO with Diphox~\cite{diphox}; Multi-leg Sherpa 2.2 generator with matrix elements for the $pp \to \gamma \gamma$ + 0,1 jet process at NLO and $pp \to \gamma \gamma$ + 2,3 jet process at LO, matched and merged with the parton shower using the MEPS@NLO prescription~\cite{mepsatNLO}.
Differential cross sections are measured as functions of the main variables of the photon pair system, together with the transverse component of $p_\mathrm{T,\gamma\gamma}$ with respect to the thrust axis, $a_\mathrm{T}$. Figure~\ref{fig:ptgammagamma}~and~\ref{fig:diphotonmass} show the differential cross section for the photon pair production as a function of $p_\mathrm{T}^{\gamma,1}$ and $m_{\gamma \gamma}$, respectively. Only the NNLO, as implemented by NNLOJET, and Sherpa NNLO predictions provide a satisfactory description of the data. The fixed-order NNLO calculation has a higher formal precision than Sherpa and leads to lower theoretical uncertainties in the perturbative QCD governed regions. Fixed-order predictions of DIPHOX and NNLOJET are not expected to be valid in regions dominated by multiple collinear and soft QCD emissions. The $m_{\gamma\gamma}$ distribution is governed by the $p_\mathrm{T}$ requirements on the individual photons of 40 GeV and 30 GeV, respectively. The $\gamma \gamma$+multi-jet configurations are dominant in the $m_{\gamma\gamma}$  < 70 GeV region. The integrated fiducial cross sections are shown in Figure~\ref{fig:integratedxsec}. NNLOJet and Sherpa MEPS@NLO are in better agreement with respect to the other predictions considered.

\section{Conclusion}
In this contribution, high-precision measurements involving the production of photons and jets at the centre-of-mass energy of 13 TeV are presented. All the measurements are in good agreement with the pQCD predictions within the theoretical uncertainties. The Sherpa multi-leg merged predictions lead to larger uncertainties, but they are in better agreement with the data in all regions. These studies will provide valuable physics inputs to PDF fits and MC generators.



\bibliography{SciPost_Example_BiBTeX_File.bib}

\begin{thebibliography}{99}
\bibitem{DirectFrag} T. Pietrycki and A. Szczurek, {\it Photon-jet correlations in $pp$ and $p\bar{p}$ collisions}, PRD 76 (2007) 034003, \doi{10.1103/PhysRevD.76.034003}.
\bibitem{gamma2jets} ATLAS collaboration, {\it Measurement of isolated-photon plus two-jet production in pp collisions at $\sqrt{s}$ = 13 TeV with the ATLAS detector}, JHEP 03 (2020) 179, \doi{10.1007/JHEP03(2020)179}.
\bibitem{diphoton} ATLAS collaboration, {\it Measurement of the production cross section of pairs of isolated photons in $pp$ collisions at 13 TeV with the ATLAS detector}, JHEP 11 (2021) 169, \doi{10.1007/JHEP11(2021)169}.
\bibitem{ATLAS} ATLAS collaboration, {\it The ATLAS Experiment at the CERN Large Hadron Collider}, JINST 3 (2008) S08033, \doi{10.1088/1748-0221/3/08/S08003}.
\bibitem{Sherpa} E. Bothmann et al., {\it Event generation with Sherpa 2.2}, SciPost Phys. 7 (2019) 034, \doi{10.21468/SciPostPhys.7.3.034}.
\bibitem{Pythia} T. Sjöstrand, S. Mrenna and P. Skands, {\it A brief introduction to PYTHIA 8.1}, Comput. Phys. Commun. 178 (2008) 852, \doi{10.1016/j.cpc.2008.01.036}.
\bibitem{nnlojet} T. Gehrmann et al., {\it jet cross sections and transverse momentum distributions with NNLOJET}, PoS RADCOR2017 (2018) 074, \doi{10.48550/arXiv.1801.06415}. 
\bibitem{diphox} T. Binoth, J. P. Guillet, E. Pilon and M. Werlen, {\it A full next-to-leading order study of direct photon pair production in hadronic collisions}, EPJC 16 (2000) 311, \doi{10.1007/s100520050024}.
\bibitem{mepsatNLO} S. Hoeche et al, {\it A critical appraisal of NLO+PS matching methods},  JHEP09(2012)049, \doi{10.1007/JHEP09\%282012\%29049}.

\end{thebibliography}

\nolinenumbers

\end{document}